\pdfoutput=1
\documentclass[sigconf,natbib=true]{acmart}

\AtBeginDocument{
  }

\usepackage{booktabs}
\usepackage{graphicx}
\usepackage{adjustbox}
\usepackage{multirow}
\usepackage[]{mdframed}
\usepackage{subcaption}

\setcopyright{none}
\renewcommand\footnotetextcopyrightpermission[1]{}

\begin{document}

\title{
Teaching Dense Retrieval Models to Specialize with Listwise Distillation and LLM Data Augmentation}

\author{Manveer Singh Tamber}
\affiliation{
  \institution{University of Waterloo}
  \city{Waterloo}
  \state{Ontario}
  \country{Canada}
}
\email{mtamber@uwaterloo.ca}

\author{Suleman Kazi}
\affiliation{
  \institution{Vectara}
  \city{Palo Alto}
  \state{California}
  \country{USA}}
\email{suleman@vectara.com}

\author{Vivek Sourabh}
\affiliation{
  \institution{Vectara}
  \city{Palo Alto}
  \state{California}
  \country{USA}}
\email{vivek@vectara.com}

\author{Jimmy Lin}
\affiliation{
  \institution{University of Waterloo}
  \city{Waterloo}
  \state{Ontario}
  \country{Canada}
}
\email{jimmylin@uwaterloo.ca}

\begin{abstract}

While the current state-of-the-art dense retrieval models exhibit strong out-of-domain generalization, they might fail to capture nuanced domain-specific knowledge. 
In principle, fine-tuning these models for specialized retrieval tasks should yield higher effectiveness than relying on a one-size-fits-all model, but in practice, results can disappoint.
We show that standard fine-tuning methods using an InfoNCE loss can unexpectedly degrade effectiveness rather than improve it, even for domain-specific scenarios. 
This holds true even when applying widely adopted techniques such as hard-negative mining and negative de-noising.
To address this, we explore a training strategy that uses listwise distillation from a teacher cross-encoder, leveraging rich relevance signals to fine-tune the retriever.
We further explore synthetic query generation using large language models.
Through listwise distillation and training with a diverse set of queries ranging from natural user searches and factual claims to keyword-based queries, we achieve consistent effectiveness gains across multiple datasets. Our results also reveal that synthetic queries can rival human-written queries in training utility. 
However, we also identify limitations, particularly in the effectiveness of cross-encoder teachers as a bottleneck.
We release our code and scripts to encourage further research.\footnote{\url{https://github.com/manveertamber/enhancing_domain_adaptation}}

\end{abstract}

\pagestyle{empty}

\maketitle

\section{Introduction}

Embedding models have emerged as key components of search pipelines.
Dense retrieval involves embedding models mapping both queries and passages into a vector space, ensuring that relevant passages are positioned close to the corresponding queries.

The BEIR benchmark~\cite{thakur2021beir} demonstrated that embedding models for retrieval can struggle when evaluated using out-of-distribution retrieval tasks.
The work showed that while the dense retrieval models studied generally outperformed BM25 on effectiveness metrics when evaluated within the same domain they were trained on, these same models underperformed BM25 on average when considering the diverse retrieval tasks in BEIR.

While the state-of-the-art (SOTA) in embedding models has continued to improve, obtaining strong out-of-domain generalization, there may still be an interest in better specializing models for certain retrieval domains or tasks.
In this work, we focus on fine-tuning SOTA BERT-base~\cite{devlin-etal-2019-bert} embedding models that lead the MTEB leaderboard~\cite{muennighoff-etal-2023-mteb} for retrieval.
We specifically focus on BERT-base models for their computational efficiency and practicality for retrieval.
In particular, we consider BGE-base \textit{\small BAAI/bge-base-en-v1.5}~\cite{bgecpack}, GTE-base \textit{\small Alibaba-NLP/gte-base-en-v1.5}~\cite{gte} and Arctic-embed-m \textit{\small Snowflake/snowflake-arctic-embed-m-v1.5}~\cite{merrick2024arctic}.
We try to improve the effectiveness of multiple SOTA embedding models on multiple retrieval datasets, ensuring that our findings generalize across different datasets and models with their varied training regimens.
We also experiment with the unsupervised variant of E5-base \textit{\small intfloat/e5-base-unsupervised}~\cite{wang2022text} to investigate fine-tuning an embedding model with strong contrastive pre-training for retrieval, but without supervised fine-tuning.
The unsupervised E5 model also allows us to study fine-tuning a model with a lower threat of task contamination, which we argue is prevalent within current SOTA embedding models that aim to be presented competitively in benchmarks such as BEIR.

Our findings reveal that, surprisingly, fine-tuning SOTA embedding models with a contrastive loss often hurts the effectiveness of the models, even when training to specialize for particular retrieval tasks, taking care to train with diverse queries, and de-noising hard negatives using a cross-encoder teacher.
However, incorporating a listwise distillation loss alongside the contrastive loss allows for effectiveness gains across a diverse range of retrieval tasks.

Enabled by the recent advances in LLMs, we also experiment with LLM data augmentation by generating a diverse set of queries in the form of natural web search queries, questions, titles, claims, and keywords, without relying on existing query examples from the retrieval dataset.
We find that training with diverse generated queries benefits retrieval across BEIR tasks, outperforming prior approaches that rely solely on synthetic user search queries or questions.
Furthermore, the quality of synthetic queries is promising, showing effectiveness that is competitive with human-written queries for training retrieval models.

While our approach achieves promising retrieval effectiveness improvements across multiple datasets and models, we also study cases where our methods were not able to improve retrieval effectiveness. These cases reveal challenges with task contamination and the need for stronger cross-encoders trained with more diverse ranking tasks to compete with advances in dense retrieval.

\section{Background}

Previous work has explored the domain adaptation of retriever models through synthetic query generation~\cite{liang2020embedding, ma2020zero}.
These approaches typically involved generating synthetic queries for many given passages and then fine-tuning retriever models using these passage-query pairs.

Alongside putting forward the BEIR benchmark, Thakur et al.~\cite{thakur2021beir} also investigated adapting retrievers to BEIR datasets using a T5-based query generator~\cite{t5, docTTTTTquery} trained on MS MARCO.
The TAS-B dense retriever\cite{tasb} fine-tuned with these queries showed mixed results, underperforming the original TAS-B and BM25 on average.
GPL~\cite{gpl} extended this work by introducing an additional step: using a cross-encoder to label query-passage pairs.
The dense retriever was then trained to mimic the score margins between positive and negative query-passage pairs from the cross-encoder.

Several more recent works have introduced improved approaches to synthetic query generation and domain adaptation.
InPars~\cite{inpars} used GPT-3~\cite{gpt3} for few-shot query generation to adapt rerankers instead of retrievers.
InPars-v2~\cite{inparsv2} later replaced GPT-3 with the open-source GPT-J~\cite{gpt-j}.
Promptagator~\cite{Promptagator} used few-shot examples of task-specific queries and passages to generate queries more in line with those from the BEIR datasets being used for evaluation.
UDAPDR~\cite{saad2023udapdr} focused on the domain adaptation of ColBERTv2~\cite{colbertv2}, a multi-vector retrieval model, leveraging GPT-3 and FLAN-T5 XXL~\cite{flant5} for query generation to train up to 10 cross-encoders that were each in turn used to annotate triples of (query, positive document, and negative document) to fine-tune retrieval models.

Unlike previous approaches, we explore the generation of diverse queries and the use of listwise distillation from cross-encoders to provide strong relevance signals for adapting retrievers.
We generate diverse queries without relying on few-shot examples from evaluation datasets to allow for fine-tuning that generalizes well to retrieval domains and tasks without needing specific human-written query examples.
We also evaluate how generated queries compare to human-written queries and we demonstrate the need for cross-encoder listwise distillation empirically, showing that a simpler contrastive learning setup fails to improve the retrieval effectiveness of current SOTA embedding models.

\section{Methodology}

\begin{table}[t]
\centering
\adjustbox{max width=0.6\columnwidth}{
\begin{tabular}{l|cc}
\toprule
\textbf{Dataset} & \textbf{BGE} & \textbf{RankT5-3B} \\
\midrule
DL19            & 70.2          & \textbf{75.1} \\
DL20            & 67.7          & \textbf{77.2} \\
TREC-COVID      & 78.1          & \textbf{85.7} \\
NFCorpus        & 37.3          & \textbf{40.4} \\
FiQA            & 40.6          & \textbf{52.6} \\
SCIDOCS         & \textbf{21.7} & 19.3 \\
ArguAna         & \textbf{63.6} & 34.9 \\
Touché-2020     & 25.7          & \textbf{38.6} \\
DBPedia         & 40.7          & \textbf{48.1} \\
FEVER           & \textbf{86.3} & 84.9 \\
Climate-FEVER   & \textbf{31.2} & 27.3 \\
SciFact         & 74.1          & \textbf{77.1} \\
\bottomrule
\end{tabular}
}
\caption{\textbf{NDCG@10} Scores for BGE and RankT5-3B reranking the top-100 BGE retrieved passages. Best scores are bolded.}
\label{tab:rerank_ndcg}
\end{table}

\subsection{Synthetic Query Generation}

Given up to 100K passages randomly sampled from a retrieval corpus, we generate queries for each passage to fine-tune the retrievers.
The assumption is that the content of the 100K passages should represent the retrieval task well and indicate what users might search for.
We generate synthetic queries using Llama-3.1 (8B)~\cite{dubey2024llama} by providing up to 3 examples of passage-query pairs and then prompting the LLM to generate a query for a given passage.
Given Wikipedia passages from BEIR's NQ corpus, we use GPT4o~\cite{achiam2023gpt} to generate high-quality queries to use as examples to guide the generation of Llama-3.1 (8B).
We examine generating six different types of queries, including questions, claims, titles, keywords, natural user search queries, and natural user search queries given human-written examples from MSMARCO~\cite{bajaj2016ms}.

Prior domain adaptation work primarily generated synthetic questions or search queries resembling human-written ones from MSMARCO~\cite{saad2023udapdr, thakur2021beir, gpl, inpars}.
Promptagator~\cite{Promptagator} focused on generating task-specific queries but did so by using up to 8 examples from the same retrieval task that the models were evaluated on.
In contrast, our methods generate diverse queries without having to rely on queries from the evaluation dataset.

For brevity, we do not provide the prompts to generate synthetic queries in this paper, but we make them available in our GitHub repository.
In general, we prompt the LLM to generate queries of under 20 words that are addressed by the given passage.
We use vLLM~\cite{kwon2023efficient} for inference and find that 100K queries for MSMARCO passages can be generated in roughly 1-2 hours using an RTX 6000 Ada GPU, although times vary depending on passage lengths.

\subsubsection{Filtering Generated Queries}
\label{sec:filtering}
To ensure the quality of generated queries, some filtering is needed.
Promptagator filtered out queries if their corresponding passage did not rank first with the unadapted retriever.
This may remove high-quality queries that are challenging for the retriever.
To improve filtering, we make use of a stronger teacher cross-encoder.
First, we discard queries where the passage isn’t in the top 20 retrieved passages.
Then, we filter out queries where the passage isn’t ranked first with the reranker.

\subsection{Evaluation}

We evaluate retrieval effectiveness after fine-tuning retrievers to adapt to BEIR~\cite{thakur2021beir} corpora and the MSMARCO~\cite{bajaj2016ms} passage ranking corpus.
We report NDCG@10 and Recall@100 scores to evaluate the fine-tuned retrievers.
For BEIR datasets, we focus on the TREC-COVID~\cite{voorhees2021trec}, NFcorpus~\cite{NFCorpus}, FiQA~\cite{fiqa}, SCIDOCS~\cite{scidocs}, ArguAna~\cite{arguana}, Touché-2020~\cite{touche}, DBPedia~\cite{dbpedia}, FEVER~\cite{fever}, CLIMATE-FEVER~\cite{diggelmann2020climate}, and SciFact~\cite{wadden-etal-2020-fact} datasets.
We choose these datasets because their corpora are under open licenses and they represent a diversity of retrieval tasks offering varying types of queries (e.g., factual claims, opinion-based questions), corpora (e.g., Wiki\-pedia, scientific abstracts, forum posts), and topics (e.g., financial, COVID-19, climate-change).
We also consider MSMARCO passage ranking to examine our fine-tuning methods in-domain.
For MSMARCO passages, we sample 200K passages to generate queries instead of 100K to allow for greater possible effectiveness boosts because MSMARCO passage ranking is a more general retrieval task that should already be in-domain for retrieval models.
After fine-tuning with MSMARCO passages, we evaluate using the TREC Deep Learning Tracks from 2019 (DL19) and 2020 (DL20)~\cite{craswell2020overview,craswell2021overview}.

\subsubsection{Prior Fine-Tuning}
\label{sec:prior_tuning}

Table~\ref{tab:rerank_ndcg} presents NDCG@10 scores across the studied datasets after reranking BGE-retrieved passages with RankT5-3B.
We focus on RankT5-3B~\cite{zhuang2022rankt5} as the reranker due to its strong effectiveness and generalization to out-of-domain tasks~\cite{zhuang2022rankt5, qin2023large, tamber2023scaling,listt5}.
While reranking generally improves effectiveness, supporting the motivation for distilling rerankers into retrievers, we observe score drops on FEVER, Climate-FEVER, ArguAna, and SCIDOCS after reranking.
This is unexpected given rerankers’ typical advantages over retrievers such as the ability to judge the passage relevance directly with respect to the query and RankT5-3B’s significantly larger size (3B parameters) compared to the BGE retriever (110M parameters).
However, the BGE model and the embedding models examined have undergone extensive fine-tuning.
Notably, while RankT5 is trained only on MSMARCO and NQ~\cite{nq}, GTE~\cite{gte} and Arctic~\cite{merrick2024arctic} include FEVER and other BEIR datasets in their training data, along with pretraining that involves scientific abstracts with corresponding titles as queries, potentially inflating their performance on SCIDOCS and similar datasets.
The BGE model's training data is not clearly mentioned~\cite{bgecpack}, but we suspect it is similarly suited for BEIR evaluation.
Table~\ref{tab:rerank_ndcg} both highlights the promise of cross-encoder distillation but also reveals challenges in fair evaluation due to potential task contamination and data leakage for retrievers being trained on BEIR datasets, potentially to remain competitive for BEIR and MTEB evaluation.

\subsubsection{Passage De-duplication} 
Similar to other previous work~\cite{pradeep2024ragnarokreusableragframework}, we identify many near-duplicate passages in MSMARCO and BEIR corpora, which may reduce generated query diversity and hinder contrastive learning if duplicates appear in training batches.
To address this, we normalize text by considering whitespace, case, and punctuation, and then remove passages that are substrings of others.
This approach eliminated over 950K from MSMARCO’s total 8.8M passages.

\subsection{Model Training}

\subsubsection{Cross-Encoder Teacher}

We use RankT5-3B~\cite{zhuang2022rankt5} as a cross-encoder teacher.
We first retrieve 20 passages for each generated query using the retriever to be fine-tuned.
We then rerank these 20 passages for the query using RankT5-3B, verifying that the corresponding passage ranks first as described in Section~\ref{sec:filtering}.
We normalize scores across all queries using min-max normalization, but using the 1st and 99th percentiles with clipping to scale data to [0,1].
The time for reranking varies depending on passage lengths, however, reranking 20 MSMARCO passages each for 100K queries takes roughly 5 hours using an RTX 6000 Ada GPU.

\subsubsection{Cross-encoder Listwise Distillation}

RocketQAv2~\cite{rocketqav2} proposed jointly training a dense retriever and a cross-encoder, allowing both models to learn from each other.
We use a similar formulation in our work to distill a cross-encoder to retrievers given a relevance distribution:

\begin{equation}
\tilde s(q_i,p_i) = {\frac{e ^ {\frac{s(q_i,p_i)}{\tau}}}{e ^ {\frac{s(q_i,p_i)}{\tau}} + \sum_{k=1}^K e ^ {\frac{s(q_i,p_{ik})}{\tau}}}}
\end{equation}

\noindent based on a scoring function $s_{de}(q,p)$ for the dense retriever, which is the cosine similarity of the query and passage embeddings and $s_{ce}(q,p)$ for the cross-encoder, which is the relevance score from the cross-encoder for a passage with respect to a query.
We minimize the KL divergence of the two relevance distributions $\tilde s_{de}(q,p)$ and $\tilde s_{ce}(q,p)$ over all the $K=19$ hard negative passages and the relevant passage for each query.
We find that the temperature parameters of $\tau=0.05$ for $\tilde s_{de}(q,p)$ and $\tau=0.3$ for $\tilde s_{ce}(q,p)$ work well for distilling into the retriever models. We tuned these hyperparameters by training BGE on MSMARCO and evaluating on DL19 and DL20. We use these values for all experiments.

\subsubsection{Positive-Aware Contrastive Loss}

\begin{figure}[ht]
    \centering
    \includegraphics[width=0.95\columnwidth]{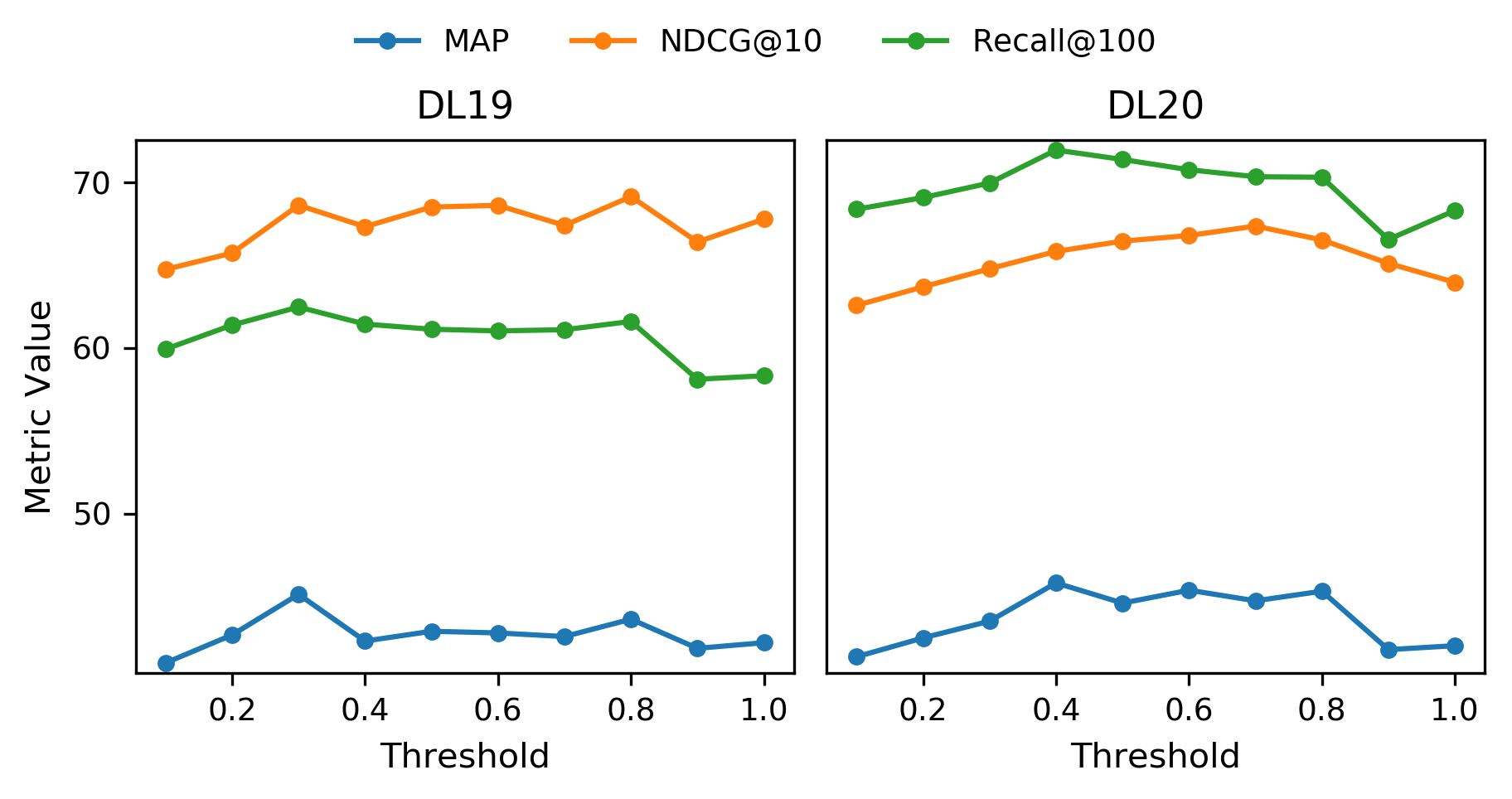} 
    \caption{Retrieval effectiveness scores on DL19 and DL20 at different hard negative filtering thresholds during BGE-base fine-tuning on MSMARCO passages.}
    \label{fig:threshold_tuning}
\end{figure}

\begin{table*}[ht!]
\centering
\scalebox{0.8}{
\begin{tabular}{l|cc|cc|cc|c}
\toprule
\textbf{Dataset} 
& \multicolumn{2}{c|}{\textbf{BGE}} 
& \multicolumn{2}{c|}{\textbf{GTE}} 
& \multicolumn{2}{c|}{\textbf{Arctic}} 
& \textbf{Promptagator} \\
& nDCG & Recall & nDCG & Recall & nDCG & Recall & nDCG \\
\midrule
DL19          
& $70.2 \;\rightarrow\; \mathbf{71.8}$ 
& $60.9 \;\rightarrow\; \mathbf{63.0}$ 
& $\mathbf{71.9} \;\rightarrow\; 71.8$ 
& $62.1 \;\rightarrow\; \mathbf{65.6}$ 
& $\mathbf{74.4} \;\rightarrow\; 74.3$ 
& $64.7 \;\rightarrow\; \mathbf{67.2}$ 
& -- \\
DL20          
& $67.7 \;\rightarrow\; \mathbf{69.7}$ 
& $71.5 \;\rightarrow\; \mathbf{75.6}$ 
& $71.5 \;\rightarrow\; \mathbf{72.6}$ 
& $69.8 \;\rightarrow\; \mathbf{72.3}$ 
& $72.1 \;\rightarrow\; \mathbf{73.6}$ 
& $74.2 \;\rightarrow\; \mathbf{75.4}$ 
& -- \\
TREC-COVID    
& $78.1 \;\rightarrow\; \mathbf{82.2}$ 
& $14.1 \;\rightarrow\; \mathbf{15.8}$
& $75.3 \;\rightarrow\; \mathbf{78.2}$
& $14.0 \;\rightarrow\; \mathbf{14.8}$
& $\mathbf{82.2} \;\rightarrow\; 80.9$
& $14.8 \;\rightarrow\; \mathbf{14.9}$
& 75.6 \\
NFCorpus      
& $37.3 \;\rightarrow\; \mathbf{38.3}$ 
& $33.7 \;\rightarrow\; \mathbf{35.8}$ 
& $35.3 \;\rightarrow\; \mathbf{37.9}$ 
& $33.1 \;\rightarrow\; \mathbf{34.6}$ 
& $36.1 \;\rightarrow\; \mathbf{37.4}$ 
& $32.4 \;\rightarrow\; \mathbf{33.7}$ 
& 33.4 \\
FiQA          
& $40.6 \;\rightarrow\; \mathbf{44.3}$ 
& $74.2 \;\rightarrow\; \mathbf{75.8}$ 
& $48.7 \;\rightarrow\; \mathbf{49.5}$ 
& $\mathbf{81.7} \;\rightarrow\; 81.4$ 
& $42.5 \;\rightarrow\; \mathbf{45.5}$ 
& $74.8 \;\rightarrow\; \mathbf{75.9}$ 
& 46.2 \\
ArguAna       
& $\mathbf{63.6} \;\rightarrow\; 61.4$ 
& $\mathbf{99.2} \;\rightarrow\; \mathbf{99.2}$ 
& $\mathbf{62.1} \;\rightarrow\; 60.9$ 
& $99.2 \;\rightarrow\; \mathbf{99.4}$ 
& $56.5 \;\rightarrow\; \mathbf{58.6}$ 
& $98.4 \;\rightarrow\; \mathbf{99.2}$ 
& 59.4 \\
Touch{\'e}-2020   
& $25.7 \;\rightarrow\; \mathbf{35.3}$ 
& $48.7 \;\rightarrow\; \mathbf{50.6}$
& $27.5 \;\rightarrow\; \mathbf{32.1}$
& $48.3 \;\rightarrow\; \mathbf{49.0}$
& $33.2 \;\rightarrow\; \mathbf{37.9}$
& $50.0 \;\rightarrow\; \mathbf{52.4}$
& 34.5 \\
DBPedia       
& $40.7 \;\rightarrow\; \mathbf{45.5}$ 
& $53.0 \;\rightarrow\; \mathbf{56.3}$ 
& $36.9 \;\rightarrow\; \mathbf{44.7}$ 
& $46.9 \;\rightarrow\; \mathbf{55.2}$ 
& $44.7 \;\rightarrow\; \mathbf{45.1}$ 
& $\mathbf{58.7} \;\rightarrow\; 57.2$ 
& 38.0 \\
SCIDOCS       
& $\mathbf{21.7} \;\rightarrow\; 19.4$ 
& $\mathbf{49.6} \;\rightarrow\; 45.5$ 
& $\mathbf{21.7} \;\rightarrow\; 20.2$ 
& $\mathbf{50.1} \;\rightarrow\; 46.6$ 
& $\mathbf{20.0} \;\rightarrow\; 18.6$ 
& $42.3 \;\rightarrow\; \mathbf{43.5}$
& 18.4 \\
FEVER         
& $\mathbf{86.3} \;\rightarrow\; 80.0$ 
& $\mathbf{97.2} \;\rightarrow\; 95.7$ 
& $\mathbf{92.1} \;\rightarrow\; 82.4$ 
& $\mathbf{97.5} \;\rightarrow\; 96.0$ 
& $\mathbf{85.6} \;\rightarrow\; 80.4$ 
& $\mathbf{97.6} \;\rightarrow\; 96.1$ 
& 77.0 \\
Climate-FEVER 
& $\mathbf{31.2} \;\rightarrow\; 25.1$ 
& $\mathbf{63.6} \;\rightarrow\; 59.3$ 
& $\mathbf{40.1} \;\rightarrow\; 30.3$ 
& $\mathbf{71.7} \;\rightarrow\; 63.8$ 
& $\mathbf{34.7} \;\rightarrow\; 27.2$ 
& $\mathbf{66.7} \;\rightarrow\; 63.1$ 
& 16.8 \\
SciFact       
& $74.1 \;\rightarrow\; \mathbf{76.2}$ 
& $96.7 \;\rightarrow\; \mathbf{97.0}$ 
& $75.5 \;\rightarrow\; \mathbf{76.3}$ 
& $97.3 \;\rightarrow\; \mathbf{97.7}$ 
& $70.5 \;\rightarrow\; \mathbf{73.6}$ 
& $94.8 \;\rightarrow\; \mathbf{95.7}$ 
& 65.0 \\
\bottomrule
\end{tabular}
}
\caption{nDCG@10 and Recall@100 for the models (before fine-tuning 
$\rightarrow$ after fine-tuning). Bolded values indicate the best score for each model on each dataset. Promptagator scores 
(nDCG only) are shown on the right for reference.}
\label{tab:full_training}
\end{table*}

\begin{table}[t]
    \centering
    \adjustbox{max width=\columnwidth}{
    \begin{tabular}{l|cccc|cc|cc}
        \toprule
         & \multicolumn{4}{c}{MSMARCO} & \multicolumn{2}{c}{SciFact} & \multicolumn{2}{c}{FiQA} \\
        \cmidrule(lr){2-5} \cmidrule(lr){6-7} \cmidrule(lr){8-9}
         & \multicolumn{2}{c}{DL19} & \multicolumn{2}{c}{DL20} &  &  &  &  \\
        \cmidrule(lr){2-3} \cmidrule(lr){4-5}
         & NDCG & Recall & NDCG & Recall & NDCG & Recall & NDCG & Recall \\
        \midrule
        BGE & 70.2 & 60.9 & 67.7 & 71.5$\dagger$ & 74.1 & 96.7 & 40.6$\dagger$ & 74.2 \\
        + FT with contrastive loss & 68.6 & 61.0 & 66.8 & 70.8 & 72.1 & 96.0 & 39.5 & 70.7 \\
        + FT with listwise loss & \textbf{71.8} & \textbf{63.6} & \textbf{69.7} & 75.1 & \textbf{76.3} & \textbf{97.0} & \textbf{44.8} & \textbf{75.9}  \\
        + FT with combined loss & \textbf{71.8} & 63.0 & \textbf{69.7} & \textbf{75.5} & 76.2 & \textbf{97.0} & 44.3 & 75.8  \\
        \midrule
        GTE & \textbf{71.9} & 62.1$\dagger$ & 71.5 & 69.8$\dagger$ & 75.5 & 97.3 & 48.7 & \textbf{81.7} \\
        + FT with contrastive loss & 71.5 & 62.6 & 67.8 & 69.8 & 71.2 & 95.5 & 45.1 & 77.1 \\
        + FT with listwise loss & \textbf{72.3} & 64.7 & \textbf{72.6} & \textbf{73.8} & 75.4 & \textbf{97.7} & 48.5 & 80.2 \\
        + FT with combined loss & 71.8 & \textbf{65.6} & \textbf{72.6} & 73.3 & \textbf{76.3} & \textbf{97.7} & \textbf{49.5} & 81.4 \\
        \midrule
        Arctic & \textbf{74.4} & 64.7 & 72.1 & 74.2 & 70.5$\dagger$ & 94.8 & 42.5$\dagger$ & 74.8 \\
        + FT with contrastive loss & 72.2 & 64.3 & 71.9 & 73.1 & 69.1 & 94.9 & 41.8 & 73.0 \\
        + FT with listwise loss & \textbf{74.4} & \textbf{67.4} & \textbf{73.6} & 75.0 & 73.2 & \textbf{96.0} & 45.2 & \textbf{75.9} \\
        + FT with combined loss & 74.3 & 67.2 & \textbf{73.6} & \textbf{75.4} & \textbf{73.6} & 95.7 & \textbf{45.5} & \textbf{75.9} \\
        \midrule
        E5-unsupervised & 56.3$\dagger$ & 52.6$\dagger$ & 54.6$\dagger$ & 62.1$\dagger$ & 74.3 & \textbf{98.7} & 40.1$\dagger$ & 71.8$\dagger$ \\
        + FT with contrastive loss & 67.9 & 61.7 & 68.2 & 71.7 & 72.3 & 95.0 & 40.5 & 73.8 \\
        + FT with listwise loss & 69.9 & \textbf{63.7} & 72.7 & 74.3 & 76.2 & 96.7 & 44.0 & 75.5 \\
        + FT with combined loss & \textbf{72.4} & 63.3 & \textbf{73.2} & \textbf{75.2} & \textbf{76.3} & 97.0 & \textbf{45.4} & \textbf{77.0} \\
        \bottomrule
    \end{tabular}
    }
    \caption{NDCG@10 and Recall@100 for the original and fine-tuned (FT) models, examining the effect of the training loss. A $\dagger$ indicates a significant difference between the original and FT model with the combined loss, based on a one-sided, paired t-test $(p < 0.05)$, with Holm--Bonferroni correction across all datasets and metrics for each model's results. Best scores for each model are bolded.}
    \label{tab:loss_results}
\end{table}

Much of the recent work in dense retrieval models, and all of the models considered in this work~\cite{bgecpack,gte,merrick2024arctic,wang2022text}, involve training the models with an InfoNCE~\cite{oord2018representation} contrastive loss that takes advantage of in-batch or mined hard- negatives to learn to represent text with embeddings contrastively.
Many of these works have also argued the importance of training with hard-negatives to train effective dense retrieval models~\cite{gte,merrick2024arctic}.
In our work, we contrast with every hard-negative passage in the batch to allow for a larger contrastive batch size using the following loss, where $K=19$ hard negatives $p_{jk}$ are mined for each query $q_j$ and used for training.
We use a temperature of $\tau=0.01$, which is the commonly used value with the models considered~\cite{bgecpack,gte,merrick2024arctic,wang2022text}:

\begin{equation}
-\frac{1}{n}{\sum_{i=1}^n \log \frac{e ^ {\frac{s_{de}(q_i,p_i)}{\tau}}}{\sum_{j=1}^n (e ^ {\frac{s_{de}(q_i,p_j)}{\tau}} + \sum_{k=1}^K e ^ {\frac{s_{de}(q_i,p_{jk})}{\tau}})}}
\end{equation}

\noindent A recent study explored techniques for mining hard negatives while acknowledging that some mined ``negatives'' may be relevant despite lacking relevance labels, which hurts contrastive learning~\cite{nvretriever}.
It found that the best method for filtering false negatives is to exclude passages with a relevance score above a certain percentage of the relevant passage's score, using a teacher embedding model.
Specifically, passages scoring above 95\% of the relevant passage’s score were removed.

In our work, we tune this threshold using the cross-encoder’s normalized scores.
Figure~\ref{fig:threshold_tuning} shows MAP, NDCG@10, and Recall@100 scores on DL19 and DL20 by the threshold applied for the BGE-base model trained using MSMARCO passages and the contrastive loss described above. We find that a threshold of 60\% works well and is a reasonable moderate threshold that balances the retrieval metrics.

\subsubsection{Training Setup}

We train models using the sum of the contrastive loss and the listwise distillation loss, finding that a combination works well with a $0.1$ weight on the contrastive loss.
This weight was tuned based on dev losses when fine-tuning BGE on MSMARCO passages.
We examine the choice of loss function further in Section~\ref{sec:eval_effectiveness}.
We train our models on a single 48GB RTX 6000 Ada or 48GB L40S GPU depending on availability.
Models were trained with a learning rate of $2e-4$ and 4096 queries per batch, leveraging GradCache~\cite{gao2021scaling} to support large contrastive batch sizes following findings that larger batch sizes help train more effective embedding models for retrieval~\cite{bgecpack, gte, merrick2024arctic}.
We use a 90\%/10\% train/dev split and train for up to 30 epochs, selecting the model with the best dev loss and stopping if the dev loss fails to improve for two epochs.
Training times vary by dataset; for MSMARCO passages, training took 48-60 hours for all models considered.

\section{Results}

\subsection{Evaluating the Effectiveness of our Methods}
\label{sec:eval_effectiveness}

Table~\ref{tab:loss_results} first shows that training with a contrastive loss underperforms training with a listwise loss and generally hurts effectiveness over the base models, except for when fine-tuning the unsupervised E5 model.
Nonetheless, training with a listwise distillation loss consistently improves effectiveness over the base models.
While combining the listwise loss and contrastive loss does not generally clearly beat training with a listwise loss alone, it does do so for E5-unsupervised which suggests that it is most helpful to incorporate a contrastive loss when the model has not already undergone strong contrastive fine-tuning.
For further experiments, we train with the combined loss as it provides arguably the strongest baseline.

Table~\ref{tab:full_training} shows that both NDCG@10 and Recall@100 consistently improve across all three models for most datasets.
However, exceptions include FEVER, Climate-FEVER, SCIDOCS, and ArguAna, where Section~\ref{sec:prior_tuning} highlighted challenges with the reranker's relative effectiveness on these particular datasets, potentially because of the prior fine-tuning of the retriever models on these datasets or retrieval tasks, unlike the reranker.
Additionally, some scores marginally decrease after fine-tuning, such as DL19 NDCG for the GTE and Arctic models, FiQA recall for the GTE model, and DBPedia recall and TREC-COVID NDCG for the Arctic model.
These results present the difficulty of performing and evaluating domain adaptation.
The Table also provides scores from Promptagator, which is the most recent domain adaptation work for single-vector dense retrievers.
Promptagator's scores are generally lower than the scores for the base models even without further fine-tuning, suggesting that Promptagator domain adaptation methods have quickly been undermined by the SOTA models.

\subsection{Evaluating the Quality of Generated Queries}

\subsubsection{Evaluating the Role of Query Types}

Table~\ref{tab:query_type} shows that while training with the query type that most aligns with the retrieval task is helpful, training with a diverse set of queries generally results in the best effectiveness and is the preferred approach.
In SciFact evaluation, where queries are scientific claims, training with MSMARCO-style user queries is most beneficial, followed by training with claims.
In MSMARCO DL19 and DL20 evaluation, training with MSMARCO-style user queries scores best on DL20, but training with titles scores marginally better on DL19.
Queries in FiQA are user-written questions and training with MSMARCO-style queries scores best.
Generally, training with all queries allows for the strongest scores, regardless of the particular query type that most aligns with the retrieval task.

\begin{table}[t]
    \centering
    \adjustbox{max width=\columnwidth}{
    \begin{tabular}{l c c c | c c c c}
        \toprule
        \textbf{Query Type} 
            & \multicolumn{3}{c}{\textbf{\#Q}} 
            & \multicolumn{4}{c}{\textbf{NDCG@10}} \\
        \cmidrule(lr){2-4} \cmidrule(lr){5-8}
        & (DL19+20) & FiQA & Sci
        & DL19 & DL20 & FiQA & SciFact \\
        \midrule
        (No fine-tuning)
            & -- & -- & --
            & 56.3 & 54.6 & 40.1 & 74.4 \\
        \midrule
        {MSMARCO User Queries} 
            & 96K & 39k & 5k 
            & 72.8 & 72.8 & 43.7 & 75.2 \\
        {Claims} 
            & 136k & 46k & 5k 
            & 65.3 & 68.0 & 40.7 & 74.5 \\
        {Titles} 
            & 99k & 34k & 5k 
            & \textbf{72.9} & 71.6 & 42.9 & 72.3 \\
        {All Queries} 
            & 646k & 240k & 30k 
            & 72.4 & \textbf{73.2} & \textbf{45.4} & \textbf{76.3} \\
        \bottomrule
    \end{tabular}
    }
    \caption{Retrieval effectiveness when training the E5-unsupervised model with different query types and all generated queries. \#Q gives the number of queries for training.}
    \label{tab:query_type}
\end{table}

\subsubsection{Comparing Human-written and Synthetic Queries}

\begin{table}[t]
    \centering
    \adjustbox{max width=\columnwidth}{
    \begin{tabular}{lccccc}
        \toprule
        \multirow{2}{*}{Query Type} & \multirow{2}{*}{\# Queries} & \multicolumn{2}{c}{DL19} & \multicolumn{2}{c}{DL20} \\
        \cmidrule(lr){3-4} \cmidrule(lr){5-6}
         & & NDCG & Recall & NDCG & Recall \\
        \midrule
        (No fine-tuning) & - & 56.3 & 52.6 & 54.6 & 62.1 \\
        \midrule
        User Queries (Human)             & 56K    & 71.6 & 63.9 & \textbf{73.4} & 73.0 \\
        User Queries Few-Shot (Synthetic)& 56K    & 71.7 & 64.9 & 70.3 & 72.1 \\
        User Queries Few-Shot (Synthetic)& 96K    & 72.8 & \textbf{65.3} & 72.8 & \textbf{74.5} \\
        User Queries Zero-shot (Synthetic)        & 56K    & 70.8 & 62.2 & 70.5 & 73.1 \\
        User Queries Zero-shot (Synthetic)        & 98K    & \textbf{74.0} & 62.9 & 71.8 & 73.7 \\
        \bottomrule
    \end{tabular}
    }
    \caption{Retrieval effectiveness for the E5-unsupervised model fine-tuned with human-written and synthetic queries. For the synthetic queries, results are provided for both a subset of 56K queries to provide a fair comparison and the full query set.}
    \label{tab:query_quality}
\end{table}

Table~\ref{tab:query_quality} compares training the E5-unsupervised model using human-written queries from the MSMARCO training set to synthetic queries generated by an LLM in either a few-shot setting (given three MSMARCO passage-query examples) or a zero-shot setting (without examples).
We focus on queries for 200K randomly sampled MSMARCO passages from the training set.

Applying the filtering process from Section~\ref{sec:filtering}, we find that human-written queries are filtered at a much higher rate, leaving only 56K of the original 200K compared to 96K for the few-shot synthetic queries and 98K for the zero-shot synthetic queries.
This is notable since RankT5-3B was fine-tuned on these same human-written queries but still filtered many out, suggesting that MSMARCO's labeled passages may not always be the most relevant for their query, as argued in previous work~\cite{ArabzadehSparseLabels}.

To ensure a fair comparison, we also train with a 56K subset of synthetic queries, aligning with the human-written query set on as many passages as possible.
In this controlled setting, models trained with few-shot synthetic queries outperform those trained with human-written queries on DL19, while zero-shot queries achieve slightly higher recall on DL20.
When we use all available synthetic queries for training, scores improve further, with the few-shot synthetic queries generally surpassing the zero-shot ones except for on DL19 when considering NDCG.
Nonetheless, models trained on human-written queries retain the highest NDCG on DL20.

Overall, these findings suggest that relatively lightweight LLMs, such as Llama-3.1 (8B), can generate synthetic queries that are competitive with human-written ones in training usefulness.

\section{Conclusion}

While previous methods in domain adaptation and dense retrieval involving contrastive learning may not be sufficient to enhance the retrieval effectiveness of SOTA embedding models in particular tasks, cross-encoder distillation and diverse synthetic queries are promising, allowing for effectiveness gains across varied datasets and models.
We show that synthetic data can rival human-written queries and relevance judgements for training.
Nonetheless, we find that cross-encoder effectiveness is a limiting factor for training effective embedding models, suggesting the importance of stronger cross-encoder teachers to further strengthen dense retrievers.

\bibliographystyle{ACM-Reference-Format}
\bibliography{custom}

\end{document}